# High Q and high gradient performance of the first medium-temperature baking 1.3 GHz cryomodule

Jiyuan Zhai[1, 2, 3, 4, *], Weimin Pan[1, 2, 3, 4, *], Feisi He[1, 2, 3, 4], Rui Ge[1, 2, 3, 4], Zhenghui Mi[1, 2, 3, 4], Peng Sha[1, 2, 3, 4], Song Jin[1, 2, 3], Ruixiong Han[1, 2, 3], Qunyao Wang[1, 2, 3], Haiying Lin[1, 2, 3], Guangwei Wang[1, 2, 3], Mei Li[1, 2, 3], Minjing Sang[1, 2, 3], Liangrui Sun[1, 2, 3], Rui Ye[1, 2, 3], Tongxian Zhao[1, 2, 3, 4], Shaopeng Li[1, 2, 3, 4], Keyu Zhu[1, 2, 3], Baiqi Liu[1, 2, 3], Xiaolong Wang[1, 2, 3], Xiangchen Yang[1, 2, 3], Xiaojuan Bian[1, 2, 3], Xiangzhen Zhang[1, 2, 3], Huizhou Ma[1, 2, 3], Xuwen Dai[1, 3], Zhanjun Zhang[1, 3], Liang Zhang[1, 3], Hui Zhao[1, 3], Runbin Guo[1, 3], Zhihui Mu[1, 3], Conglai Yang[1, 3], Xiaobing Zheng[1, 3], Chao Dong[1, 2, 3], Tongming Huang[1, 2, 3, 4], Qiang Ma[1, 2, 3], Hongjuan Zheng[1, 2, 3], Ming Liu[1, 2, 3, 4], Zihan Wang[1, 2, 3, 4], Wenzhong Zhou[1, 2, 3, 4]

[1] *Institute of High Energy Physics, Chinese Academy of Sciences, Beijing 100049, People's Republic of China*
[2] *Key Laboratory of Particle Acceleration Physics & Technology, Institute of High Energy Physics, Chinese Academy of Sciences, Beijing 100049, People's Republic of China*
[3] *Center for Superconducting RF and Cryogenics, Institute of High Energy Physics, Chinese Academy of Sciences, Beijing 100049, People's Republic of China*
[4] *University of Chinese Academy of Sciences, Beijing 100049, People's Republic of China*

World's first 1.3 GHz cryomodule containing eight 9-cell superconducting radio-frequency (RF) cavities treated by medium-temperature furnace baking (mid-T bake) was developed, assembled and tested at IHEP for the Dalian Advanced Light Source (DALS) and CEPC R&D. The 9-cell cavities in the cryomodule achieved an unprecedented highest average $Q_0$ of $3.8 \times 10^{10}$ at 16 MV/m and $3.6 \times 10^{10}$ at 21 MV/m in the horizontal test. The cryomodule can operate stably up to a total CW RF voltage greater than 191 MV, with an average cavity CW accelerating gradient of more than 23 MV/m. The results significantly exceed the specifications of CEPC, DALS and the other high repetition rate free electron laser facilities (LCLS-II, LCLS-II-HE, SHINE, S$^3$FEL). There is evidence that the mid-T bake cavity may not require fast cool-down or long processing time in the cryomodule. This paper reviews the cryomodule performance and discusses some important issues in cryomodule assembly and testing.

## I. INTRODUCTION

High $Q_0$ 1.3 GHz cryomodules operating in the CW mode are the key equipment for several major scientific projects, which are under construction or being planned in the world. For example, 75 high $Q_0$ 1.3 GHz cryomodules will be built for Shanghai High Repetition Rate X-ray FEL and Extreme Light Facility (SHINE) for its 8 GeV superconducting linac [1], 26 cryomodules are needed for Shenzhen Superconducting Soft X-Ray Free Electron Laser (S$^3$FEL) to accelerate the electron beam to 2.5 GeV [2], and high repetition rate upgrade of the European XFEL will replace the existing front-end cryomodules with 17 new CW high $Q_0$ cryomodules [3]. High $Q_0$ will save a large part of the construction and operation cost of the cryogenic system for these projects, or enable higher gradient operation thus higher beam energy or a shorter linac for a given cooling capacity.

The successful commissioning and operation of Linac Coherent Light Source II (LCLS-II) at SLAC in 2023 first time demonstrated highest $Q_0$ ($2.8 \times 10^{10}$) cavity operation at an average accelerating gradient of 16 MV/m in a CW superconducting RF linac of 40 cryomodules [4] with nitrogen-doping (N-doping) technology [5]. Constrained by the tunnel length, the energy upgrade of LCLS-II, i.e. LCLS-II-HE, increases the nominal operating gradient from 16 MV/m to 20.8 MV/m while keeping the high $Q_0$ of $2.7 \times 10^{10}$. Fermilab achieved these ambitious performance goals in LCLS-II-HE prototype and production cavities and cryomodules [6, 7] by improving the nitrogen doping technology with substantial efforts. Average usable gradient of 24.4 MV/m has been achieved with average $Q_0$ of $3.0 \times 10^{10}$ at 20.8 MV/m in the five tested cryomodules.

Medium-temperature furnace baking (mid-T bake) is a novel high $Q_0$ recipe discovered in 2019 [8, 9] after the previously only medium-gradient high $Q_0$ heat treatment method of N-doping since 2013. Mid-T bake was developed by Fermilab initially for quantum applications to remove $Nb_2O_5$ layer to enable high $Q_0$ at mK and low photon counts [10]. They applied the 250 ~ 400 C in-situ mid-T bake of the assembled 1.3 GHz single-cell cavity and found high $Q_0$ and anti-Q-slope similar to N-doping. The high $Q_0$ is preserved even after oxide regrows. After the pioneering work of Fermilab, KEK simplified the implementation of the recipe by mid-T baking of the unassembled cavity in a normal vacuum furnace [11], which is more accessible than the original in-situ mid-T bake. KEK's systematic investigations showed that 300 C bake has the highest $Q_0$ [12].

* zhaijy@ihep.ac.cn, panwm@ihep.ac.cn

IHEP further simplified mid-T furnace bake procedure to only one bulk EP (without light EP) and successfully applied it to 1.3 GHz 9-cell cavities in October 2020 [13, 14]. Fourteen mid-T 9-cell cavities have been tested since 2020 with high reliability by industrial vendors. In June 2023, a 1.3 GHz cryomodule with eight mid-T furnace baked 9-cell cavities was assembled and tested at IHEP and achieved world leading high $Q_0$ and high gradient for the proposed high energy circular collider CEPC and the high repetition rate EUV-FEL Dalian Advanced Light Source (DALS) R&D. Researches from other labs [1, 15-17] also confirmed the mid-T furnace bake a reproducible and stable process for high $Q_0$ cavities.

There are some distinct advantages of mid-T bake compared to nitrogen doping [18]: Due to the absence of N gas exposure, mid-T bake prevents the formation of NbN precipitates, which become defects. In the same time, lower temperature reduces the risk of contamination. Thus no uniform few-micron post-doping light EP is needed, which is challenging for low beta elliptical cavities and complex geometries. Furthermore, no intermediate EP between high temperature baking and doping or mid-T bake is required, thus substantially simplified the cavity chemical treatment to only one bulk EP, which is even simpler than the EP baseline cavities (two EP treatment needed). Data show that the $Q_0$ and gradient performance of mid-T cavities is superior to 2/6 doping of LCLS-II cavities and comparable to 2/0 and 3/60 doping recipes for LCLS-II-HE.

With these advantages, mid-T furnace bake high $Q_0$ recipe was recently adopted for the low-beta 650 MHz 5-cell cavities for Fermilab's PIP-II project [19]. The ongoing and future high repetition rate FEL and ERL projects (e.g. SHINE, $S^3$FEL, DALS and CW upgrade of European XFEL etc.), as well as the energy frontier circular lepton colliders (CEPC, FCC-ee) are all considering to use mid-T furnace bake treatment for their large number of cavities.

Therefore, demonstration of mid-T bake cavity performance in a real cryomodule is essential for the development and practical application of this novel high $Q_0$ technology. In this paper, we report the performance of the world's first mid-T bake cryomodule and discuss the important issues during the assembly, cooling down, processing and testing.

## II. VERTICAL TEST PERFORMANCE OF THE MID-T CAVITIES

Up to now, twelve mid-T bake 1.3 GHz 9-cell cavities (N5-N16) without helium vessel have been vertically tested at IHEP (Fig.1). Average maximum gradient of the cavities is 26 MV/m. Average $Q_0$ is $4.5\times10^{10}$ at 16~23 MV/m. The best cavity reached 31.3 MV/m with $Q_0$ of $5.4\times10^{10}$ at 21 MV/m and $4.9\times10^{10}$ at 31 MV/m. Details of the surface treatment including mid-T bake of these cavities can be found in Reference [14].

Eleven of these cavities (expect N6) were welded with helium vessel and carried out vertical test again. Two cavities (N17 and N18) were only vertically tested after welding the helium vessel. $Q_0$ of some of the jacketed cavities is lower than the bare cavities because of possible thermal current and not-optimized fast cooldown. Due to the tight schedule, there is no time to do the thermal cycle (warm-up to 40 K and fast cooldown to 2 K again) to recover the degraded $Q_0$ after quenching. Three cavities (N13, N16, N18) are using the single-cavity horizontal test (HT) data with high $Q_{ext}$ antenna instead of the vertical test data, as shown in Figure 2.

All the vertical test $Q_0$ values are corrected with stainless steel flanges losses (0.8 nΩ) in order to compare with cryomodule test results directly.

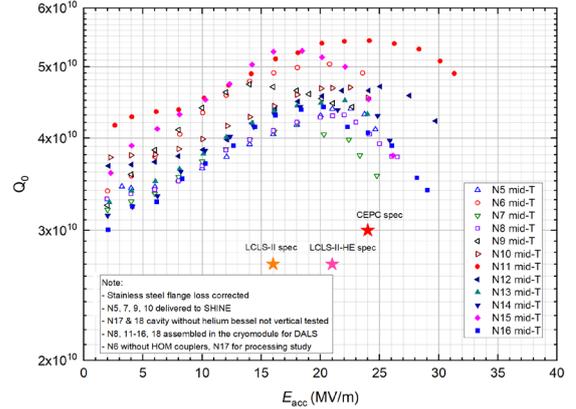

FIG. 1. Vertical test result of the 12 mid-T bake 1.3 GHz 9-cell cavities without helium vessel.

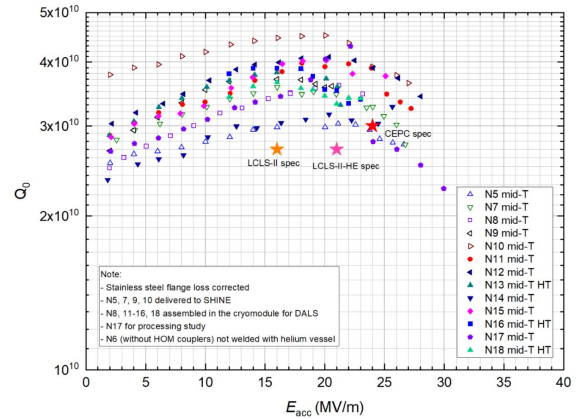

FIG. 2. Vertical test or single-cavity horizontal test (with high $Q_{ext}$ antenna) result of the mid-T bake 1.3 GHz 9-cell cavities with helium vessel. The single-cavity horizontal test was done in a special cryostat.

## III. CRYOMODULE ASSEMBLY

The cryomodule structure and assembly procedure are similar to LCLS-II cryomodules which is modified from Euro-XFEL design. The 12 m long cryomodule contains 8 1.3 GHz 9-cell cavities, 8 power input couplers, 8 tuners with piezos, one conduction-cooled superconducting magnet and one BPM. Measures to reduce microphonics were adopted according to LCLS-II experience [20]. To avoid contamination risk, the cavity string was kept in vacuum after leak check all through to the horizontal test. To reduce the multipacting processing time, the cavity string was also pumped by a turbo molecule pump whenever possible during the cold mass assembly outside the clean room for better cavity vacuum.

Temperature sensors were mounted on the top and bottom of cell #1 and cell#9 of cavities 1, 5, and 8 to monitor the temperature difference at the critical temperature. Flux gates were mounted on the outside of the cavity wall inside the helium vessel in different directions for cavities 1, 5, and 8 as well as between the two magnetic shield layers to monitor the environmental and thermal current magnetic field and flux expulsion. Six radiation detectors around the cryomodule and two Faraday cups in the upstream and downstream end of the beam pipes were used to detect the field emission and dark current.

The temperatures of all the 16 HOM coupler feedthroughs and 8 input coupler cold windows were also monitored. They were found to be critical to monitor the overheating of the HOM couplers caused by poor thermal contact. For example, in the first round assembly and testing of the mid-T cryomodule, three cavities were limited to very low gradient of only 3 MV/m, 10 MV/m and 15 MV/m. We also found overheating of the input coupler cold window of cavity#8 caused by poor mechanical connection of the warm part and cold part of the inner conductor.

## IV. MULTIPACTING PROCESSING AND GRADIENT PERFORMANCE

TESLA-shape cavities treated with high $Q_0$ recipe (e.g. N-doping, no low temperature bake after HPR) sometimes need more multipacting processing than EP baseline recipe (120 C bake after HPR), especially in the cryomodule [6, 21]. This poses a particular challenge to the relevant machines whose nominal operation gradient lies within the range of multipacting band at 17-24 MV/m of the TESLA-shape cavity.

For the mid-T cryomodule, it took two days processing to increase the gradient of the eight 9-cell cavities to their maximum gradient. Two of the cavities (CAV2 and CAV8) showed repetitive quenching during the processing, while the other cavities had nearly no multipacting quench. CAV2 took four hours to reach 19 MV/m (14 quenches at < 14 MV/m, 12 quenches at 17-19 MV/m). CAV8 took five hours to reach 26 MV/m (57 quenches at 17-24 MV/m).

CAV2 quenched at 20 MV/m due to overheating of the HOM coupler feedthrough on the field pick-up side. It is necessary to further improve the thread and bolt structure of the thermal anchor of the HOM coupler feedthroughs to avoid large gradient drop in the cryomodule, especially for long distance transportation. CAV5 had severe outgassing in the warm part of the input coupler. Pulsed RF conditioning was carried out to solve the problem.

Figure 3 shows the vertical test gradient, cryomodule maximum gradient and usable gradient for each cavity. The usable gradient is defined as the maximum gradient at which the following three conditions are met: (1) radiation level is below 0.5 mSv/h. (2) the cavity can run stably for one hour. (3) 0.5 MV/m below the quench field.

Eight cavities were powered to 16 MV/m one by one to monitor the field emission (FE) and radiation dose. CAV1 FE started at 10 MV/m and reached maximum radiation of 5 μSv/h. CAV6 FE started at 10.8 MV/m and reached maximum radiation of 13 μSv/h. The other six cavities had no FE (Fig. 4). When all eight cavities were operating at 16 MV/m, the maximum radiation dose of the six detectors was 0.08 mSv/h, which was much lower than the specification of 0.5 mSv/h. The field emission control results verified the clean assembly procedure of the cavity string as well as the end beam pipes connections.

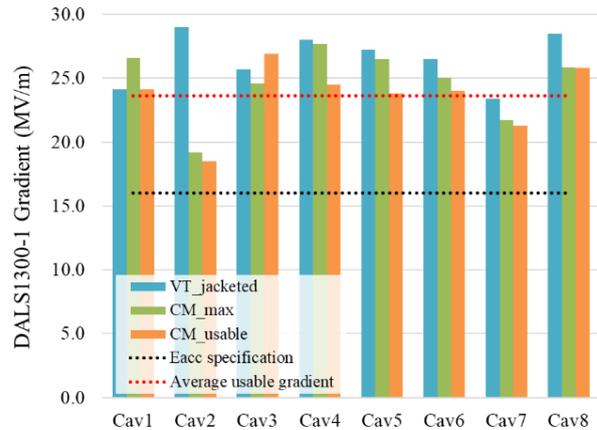

FIG. 3. Comparison of vertical test gradient, cryomodule maximum gradient and cryomodule usable gradient for each cavity in the mid-T bake cryomodule.

Dark current was measured by Faraday cups at each end of the beam pipe outside of the cryomodule when all the eight cavities operating at 16 MV/m in SELAP mode. The dark current was always smaller than the background and the specification of 1 nA by RF phase scanning.

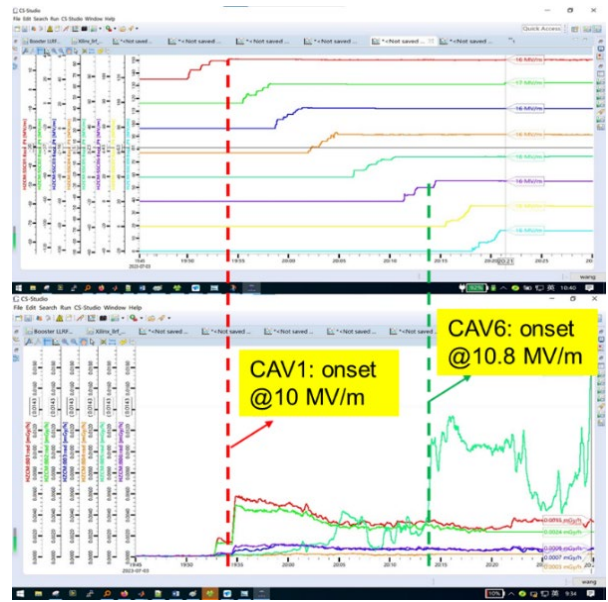

FIG. 4. Field emission onset and radiation dose of the eight cavities in the cryomodule.

The cryomodule stably reached to a total CW RF voltage of 191.2 MV when all eight cavities and the superconducting magnet were powered on. The average cavity CW gradient is 23.1 MV/m, which is higher than the average gradient of the LCLS-II cavities in the cryomodule (18.2 MV/m) [4] and comparable to the average gradient of the LCLS-II-HE cryomodules (24.5 MV/m) [7]. The cryomodule performed continuous stable operation for 12 hours at 133 MV (all cavities working at 16.0 MV/m) without any trip. The temperature of the input coupler cold window reached 90 K and was near saturation, which was lower than the specification of 100 K.

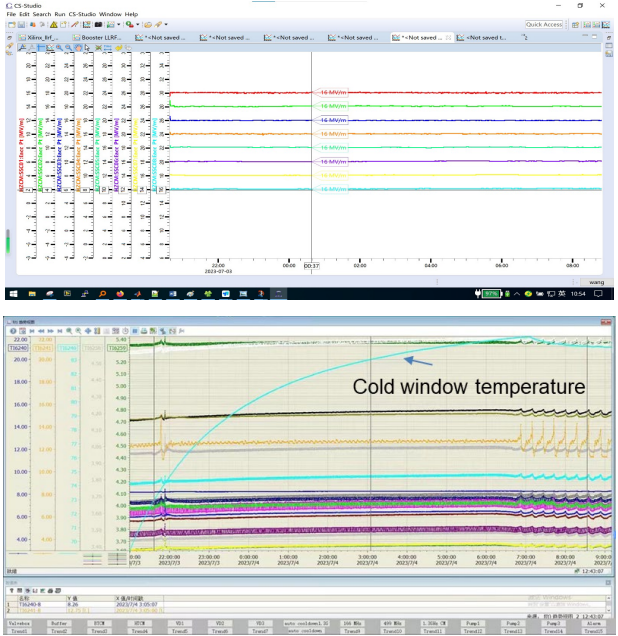

FIG. 5. Mid-T bake cryomodule stable operation at 133 MV for 12 hours. Top: gradient of the eight cavities. Bottom: input coupler cold window temperature rise.

Microphonics was measured using the RF phase deviation and converted to frequency detuning. Maximum detuning of the eight cavities was ± 4 Hz at 3 MV/m (open loop), ± 7 Hz at 8 MV/m (open loop) and ± 4 Hz at 16 MV/m (SELAP closed loop), which meet the specification of peak to peak detuning < 10 Hz.

## V. $Q_0$ PERFORMANCE

After all the cavities were processed to the maximum gradient with multiple quenches, the cryomodule was warmed up to 40 K and cooled down to 4.2 K again. The static and dynamic 2 K heat load of the cryomodule were measured using the mass flow rate method. A heater in the liquid level tank at the end of the two-phase pipe was used to calibrate the mass flow.

Total 2 K heat load of the cryomodule at 133 MV (16 MV/m average gradient) is 83.5 W, which fulfils the specification of DALS (130 W) as well as LCLS-II and SHINE (93 W). Total 2 K heat load of the cryomodule at 173 MV (21 MV/m average gradient) is 133 W, which fulfils the specification of LCLS-II-HE (137 W). The average $Q_0$ of eight cavities was calculated from the total 2 K dynamic heat load of the cryomodule. The heat load of the input couplers was very small, so it was included in the cavity dynamic heat load. The average $Q_0$ is $3.8 \times 10^{10}$ at 16 MV/m and $3.6 \times 10^{10}$ at 21 MV/m, which are far beyond the specifications of the modern CW superconducting accelerators. The static heat load of this cryomodule is 25.2 W. The $Q_0$ of the individual cavities was measured at 16 MV/m and 21 MV/m. Figure 6 shows the cavity $Q_0$ in the cryomodule compared with $Q_0$ in the vertical test.

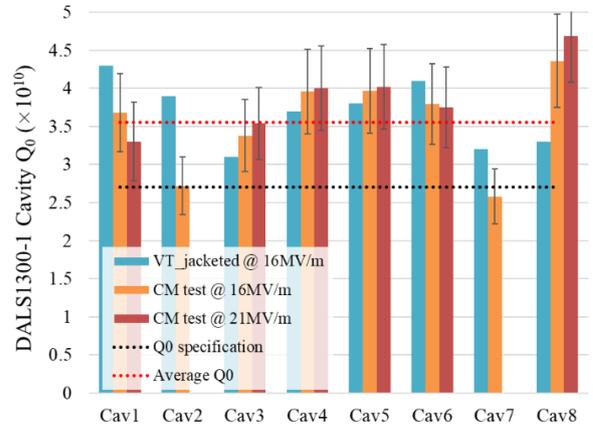

FIG. 6. Comparison of cavity $Q_0$ in vertical test at 16 MV/m, $Q_0$ in the cryomodule at 16 MV/m and 21 MV/m.

TABLE I. Summary of cavity performance in cryomodule and vertical test. The vertical test $Q_0$ values are corrected for stainless steel flange losses (0.8 n$\Omega$). $E_{max}$ and $E_{use}$ in MV/m.

| Slot in CM | CAV Serial No. | Bare Cavity Vertical Test | | | Jacketed Cavity Vertical Test | | | Cryomodule Test | | | |
|---|---|---|---|---|---|---|---|---|---|---|---|
| | | $E_{max}$ | $Q_0 / 10^{10}$ | | $E_{max}$ | $Q_0 / 10^{10}$ | | $E_{max}$ | $E_{use}$ | $Q_0 / 10^{10}$ | |
| | | | 16 MV/m | 21 MV/m | | 16 MV/m | 21 MV/m | | | 16 MV/m | 21 MV/m |
| CAV1 | N13 | 24.0 | 4.4 | 4.5 | 23.3 | 4.2 | 4.4 | 26.6 | 24.1 | 3.7 | 3.3 |
| CAV2 | N18 | / | / | / | 32.0 | 2.6 | 2.5 | 19.2 | 18.5 | 2.7 | / |
| CAV3 | N14 | 25.9 | 4.3 | 4.4 | 25.7 | 3.0 | 3.1 | 24.6 | 26.9 | 3.4 | 3.5 |
| CAV4 | N12 | 29.7 | 4.4 | 4.6 | 28.0 | 4.0 | 4.1 | 27.7 | 24.5 | 4.0 | 4.0 |
| CAV5 | N11 | 31.3 | 5.1 | 5.4 | 27.2 | 3.8 | 4.0 | 26.5 | 23.8 | 4.0 | 4.0 |
| CAV6 | N08 | 26.5 | 4.1 | 4.3 | 23.2 | 3.5 | 3.6 | 25.0 | 24.0 | 3.8 | 3.8 |
| CAV7 | N15 | 26.1 | 5.2 | 5.0 | 25.1 | 4.0 | 3.9 | 21.7 | 21.3 | 2.6 | / |
| CAV8 | N16 | 29.1 | 4.3 | 4.2 | 29.1 | 4.3 | 4.3 | 25.9 | 25.8 | 4.4 | 4.7 |
| **Average** | | **27.5** | **4.5** | **4.6** | **26.7** | **3.7** | **3.7** | **24.7** | **23.6** | **3.6** | **3.8** |

Table I summarizes the cavity performance in the cryomodule. The average gradient of the eight cavities decreases from 27.5 MV/m in the bare cavity vertical test to 26.7 MV/m in the jacketed cavity vertical test, and finally to maximum 24.7 MV/m and usable 23.6 MV/m in the cryomodule test. There is large difference of $Q_0$ between bare cavities and jacketed cavities as mention previously, while the average $Q_0$ in the cryomodule is similar to the jacketed cavity vertical test. The uncertainty in the gradient measurement is approximately 5 %. The uncertainty in the average $Q_0$ measurement from the total 2 K dynamic heat load is approximately 10 %. The uncertainty in the individual cavity $Q_0$ measurement is higher because of the very low dynamic heat load of a high $Q_0$ cavity compared to the cryogenic fluctuation.

The above heat load and cavity $Q_0$ values were measured at 2 K and the cooldown flow rate from 40 K to 4.2 K was higher than 40 g/s (because it was out of the measurement range of the flow meter, but should be less than 60 g/s). The remnant magnetic field on the cavity wall at 2 K was less than 5 mG. Preliminary studies showed that the average cavity $Q_0$ values in the cryomodule were similar for slow (~ 7 g/s) cool down with fast cool down (> 40 g/s), and the top to bottom cavity temperature difference at the critical temperature are also similar for these two cases. Further investigation on the mid-T cavities are needed regarding the relation of $Q_0$ with different cooldown rate, remnant magnetic field, thermal current and flux expulsion.

## VI. CONCLUSIONS AND OUTLOOK

World's first medium-temperature baked (mid-T) high $Q_0$ 1.3 GHz cryomodule is developed at IHEP. Main performance of the cryomodule including the total RF voltage, cryogenic heat load and radiation dose meets the requirement of DALS, S$^3$FEL and SHINE, and is beyond LCLS-II-HE and CEPC specification. The module can operate stably up to a total RF voltage greater than 191 MV with an average cavity gradient of more than 23 MV/m, and has unprecedented high average $Q_0$ of $3.8 \times 10^{10}$ at 16 MV/m and $3.6 \times 10^{10}$ at 21 MV/m. Mid-T bake is expected to be a feasible and promising technical approach for future high $Q_0$ superconducting accelerators.

More mid-T bake 1.3 GHz cavities and cryomodules will be built for China's FEL projects in next few years for further investigations and improvements, for example: (1) optimize vertical test procedure of jacketed cavities to have similar $Q_0$ with bare cavities, because bare cavities will not perform vertical test in mass production. (2) Reduce module static heat load in collaboration with the TTC member institutes. (3) Investigate multipacting processing and input coupler outgassing issues and reduce processing time. (4) Increase mid-T cavity gradient, reduce $Q_0$ spread.

## ACKNOWLEDGEMENT

The authors would like to thank the colleagues from Dalian Institute of Chemical Physics (DICP), Chinese Academy of Sciences (CAS) and Institute of Advanced Science Facilities, Shenzhen (IASF). This work was supported by the DALS, CEPC, SHINE R&D and the Platform of Advanced Photon Source (PAPS) fundings.